\documentclass[11pt]{article}
\usepackage{amsmath,amssymb,color,graphics,epsfig}

\usepackage{amsfonts}

\newcommand{\be}{\begin{equation}}
\newcommand{\ee}{\end{equation}}
\newcommand{\bea}{\setlength\arraycolsep{2pt} \begin{eqnarray}}
\newcommand{\eea}{\end{eqnarray}}

\def\ft#1#2{{\textstyle{\frac{\scriptstyle #1}{\scriptstyle #2} } }}
\def\fft#1#2{{\frac{#1}{#2}}}

\def\0{{\sst{(0)}}}
\def\1{{\sst{(1)}}}
\def\2{{\sst{(2)}}}
\def\3{{\sst{(3)}}}
\def\4{{\sst{(4)}}}
\def\5{{\sst{(5)}}}
\def\6{{\sst{(6)}}}
\def\7{{\sst{(7)}}}
\def\8{{\sst{(8)}}}
\def\sst#1{{\scriptscriptstyle #1}}

\begin{document}

\begin{flushright}
\end{flushright}

\vspace{25pt}
\begin{center}
{\large {\bf $SU(2)$-Colored (A)dS Black Holes in Conformal Gravity }}

\vspace{10pt}
Zhong-Ying Fan and H. L\"u

\vspace{10pt}

{\it Department of Physics, Beijing Normal University, Beijing 100875, China}

\vspace{40pt}

\underline{ABSTRACT}
\end{center}

We consider four-dimensional conformal gravity coupled to the $U(1)$ Maxwell and $SU(2)$ Yang-Mills fields.  We study the structure of general black hole solutions carrying five independent parameters: the mass, the electric $U(1)$ and magnetic $SU(2)$ charges, the massive spin-2 charge and the thermodynamical pressure associated with the cosmological constant, which is an integration constant in conformal gravity.  We derive the thermodynamical first law of the black holes.  We obtain some exact solutions including an extremal black hole with vanishing mass and entropy, but with non-trivial $SU(2)$ Yang-Mills charges.  We derive the remainder of the first law for this special solution.  We also reexamine the colored black holes and derive their first law in Einstein-Yang-Mills gravity with or without a cosmological constant.

\vfill {\footnotesize Emails: zhyingfan@gmail.com \ \ \ mrhonglu@gmail.com}

\thispagestyle{empty}

\pagebreak

\tableofcontents
\addtocontents{toc}{\protect\setcounter{tocdepth}{2}}




\section{Introduction}

The existence of analytical solutions is rather significant in Einstein's General Relativity which is a highly non-linear theory.  One class of solutions describe black holes and they play an important role in understanding General Relativity.  Many exact black hole solutions have been found in Einstein gravity and supergravities in diverse dimensions.  There are a few notable exceptions.  One example is the charged rotating black holes in Einstein-Maxwell theories in dimensions higher than four.  The other is the four-dimensional black hole carrying the Yang-Mills charge.  The purpose of this paper is to address the latter problem.

For asymptotically-flat Minkowski spacetimes, there exist no-hair theorems for a variety of theories.  The situation for the Yang-Mills hairs is rather subtle.  Particle-like static solution carrying Yang-Mills charges were studied in \cite{Bartnik:1988am}.  Through numerical analysis, it was demonstrated \cite{Bizon:1990sr} that colored black holes supported by the $SU(2)$ Yang-Mills fields do exist.  However, the solutions carry no global Yang-Mills charges and the numerical analysis requires a delicate fine tuning. The results were generalized to including rotation \cite{Kleihaus:2000kg,Kleihaus:2002ee}. When a cosmological constant is added into the Lagrangian, the condition for the no-hair theorem in asymptotic de Sitter or anti-de Sitter spacetime ((A)dS) is much relaxed. It was established also by numerical analysis that $SU(2)$-colored black holes with global charges also exist \cite{Torii:1995wv,Winstanley:1998sn}. The only known examples of exact solutions involving non-abelian Yang-Mills fields were constructed in supergravities, where the solutions satisfy the supersymmetric BPS conditions \cite{Chamseddine:1997nm,Meessen:2008kb,Hubscher:2008yz,Canfora:2012ap,Bueno:2014mea}.

In this paper we consider conformal gravity coupled to conformal fields such as the $U(1)$ Maxwell and the $SU(2)$ Yang-Mills fields.  Conformal gravity is an important ingredient for constructing critical gravity, where the massive spin-2 modes are replaced by some logarithmic modes \cite{Lu:2011zk}.  It was shown that with some suitable boundary conditions, Einstein gravity with a cosmological constant $\Lambda$ can emerge from conformal gravity of a certain coupling constant that is related to $\Lambda$ \cite{Maldacena:2011mk}. For spherically-symmetric Ansatz, the local conformal invariance implies that the metric can be specified by only one function instead of the usual two.  Thus, although conformal gravity contains higher derivatives, the equations of motion for the colored black hole Ansatz actually become simpler than those in the Einstein-Yang-Mills theory.  For the analogous reason, one can find the most general spherically symmetric solution in conformal gravity \cite{Riegert:1984zz}.  The vacuum solution may provide an explanation to the galactic rotation curves without introducing dark matter \cite{Mannheim:1988dj}.  Conformal gravity admits the general Kerr-AdS-NUT-like solution \cite{Mannheim:1990ya}, whose global structure as rotating black holes were analysed in \cite{Liu:2012xn}.

In this paper, the simplification in the equations of motion allows us to construct some exact $SU(2)$-colored asymptotic (A)dS black holes that carry continuous global Yang-Mills charges. (Numerical analysis of this system was performed in \cite{Brihaye:2009hf,Brihaye:2009ef}.)  For higher-derivative gravities such as conformal gravity, the mass of a black hole can be difficult to define.  Furthermore, black holes can contain also massive spin-2 hairs.  The thermodynamical first law can thus be complicated.  We shall adopt the Wald formalism \cite{wald1,wald2} to derive the first law for our colored black holes.

The paper is organized as follows.  In section 2, we consider four-dimensional gravity coupled to Yang-Mills and Maxwell fields.  We present the spherically-symmetric and static Ans\"atze for all fields and derive the equations of motion.  We study the asymptotic (A)dS black holes and find that the general solution involves five parameters.  In section 3, we present many examples of new exact solutions that carry Yang-Mills charges.  In section 4, we derive the thermodynamical first law using the Wald formalism.  In section 5, we revisit the $SU(2)$-colored black hole in Einstein-Yang-Mills theory with and without a cosmological constant, and derive the corresponding first law.  We conclude our paper in section 6.  In appendix A, we present the full set of equations of motion of $U(1)$-charged, $SU(2)$-colored solution with sphere/torus/hyperbolic symmetries.

\section{The theory and the black holes}

\label{theory}

\subsection{Conformal gravity with Yang-Mills and Maxwell fields}

In this paper, we consider conformal gravity in four dimensions, which can be built from the square of the Weyl tensor.  In four dimensions, Maxwell or general non-abelian Yang-Mills fields are also conformal.  Thus the conformal invariance is preserved when conformal gravity is minimally coupled to these gauge fields.  The Lagrangian is
\be
\mathcal{L}=\alpha\sqrt{-g}\Big(\fft12 C_{\mu\nu\rho\sigma}C^{\mu\nu\rho\sigma}-\frac{1}{2 g_s^2}\ F^a_{\mu\nu}F^{a\mu\nu}-\frac{1}{2 e^2}\ \mathcal{F}_{\mu\nu}\mathcal{F}^{\mu\nu} \Big) \label{genlag}\,,
\ee
where $C^{\mu\nu\rho\sigma}$ is the Weyl tensor and $F^a$ and ${\cal F}$ are the field strengths of the $SU(2)$ Yang-Mills and $U(1)$ gauge fields
\be
F^a_{\mu\nu}=\partial_\mu A^a_\nu-\partial_\nu A^a_\mu+\epsilon^{abc}A^b_\mu A^c_\nu\,,\qquad \mathcal{F}_{\mu\nu}=\partial_\mu \mathcal{A}_\nu-\partial_\nu \mathcal{A}_\mu\,.\label{ymdef}
\ee
The Yang-Mills and Maxwell equations of motion are
\be \partial_\mu(\sqrt{-g}F^{a\mu\nu})+\sqrt{-g}\epsilon^{abc}A_\mu^b F^{c\mu\nu}=0\,, \qquad \partial_\mu (\sqrt{-g}\mathcal{F}^{\mu\nu})=0\,.
\ee
The Einstein equations of motion are
\bea
&&-\alpha(2\bigtriangledown^{\rho}\bigtriangledown^{\sigma}+R^{\rho\sigma})
C_{\mu\rho\sigma\nu}-\frac{1}{2g_s^2}
\alpha(g^{\rho\sigma}F^a_{\mu\rho}F^a_{\nu\sigma}-\frac 14 F^2 g_{\mu\nu})\cr
&&\qquad\qquad\qquad\qquad\qquad\qquad-\frac{1}{2e^2}\alpha
(g^{\rho\sigma}\mathcal{F}_{\mu\rho}\mathcal{F}_{\nu\sigma}-\frac 14 \mathcal{F}^2 g_{\mu\nu})=0\,,
\eea
where $F^2\equiv F^a_{\mu\nu}F^{a\mu\nu}$, and $\mathcal{F}^2\equiv \mathcal{F}_{\mu\nu}\mathcal{F}^{\mu\nu}$.

The sign of each term in the Lagrangian (\ref{genlag}) is chosen with following considerations.  Higher derivative theories contain in general ghost-like modes and conformal gravity is no exception.  For (\ref{genlag}) with a positive $\alpha$ constant, the massive spin-2 modes are ghost-like whilst the massless graviton modes have the positive kinetic energy. Thus for our choice of parameters, the massive spin-2 modes are the only ghosty ones when $\alpha >0$ and $e^2>0$.

It is worth remarking that the linearity of the Maxwell equations implies that ${\cal A}$ can be scaled arbitrary which has the effect of modifying its coupling constant $e$. It follows that one can set $e^2=\pm 1$ without loss of generality.  This is no longer true for the non-linear Yang-Mills fields that are defined in this paper by (\ref{ymdef}), which contains no $g_s$.  When a solution involves a specific $g_s$, it is determined by the equations of motion, rather than by a conventional choice.

\subsection{Black hole Ansatz}

We consider static and spherically symmetric ansatz for the metric. Up to an overall scaling factor, the most general Ansatz for the metric takes the form \cite{Riegert:1984zz}
\be
ds^2 = - f(r) dt^2 + \fft{dr^2}{f(r)} + r^2 (d\theta^2 + \sin^2\theta d\varphi^2)\,.
\label{metans1}
\ee
This is in fact not the simplest Ansatz for spherical symmetry.  The metric is conformally locally equivalent to
\be
ds^2 = - h(\rho) dt^2 + \fft{d\rho^2}{h(\rho)} + d\theta^2 + \sin^2\theta d\varphi^2\,.
\label{metans2}
\ee
It is easy to see that one can first make a coordinate transformation $\rho=1/r$ and then multiply the metric (\ref{metans2}) by conformal factor $r^2$. The resulting metric (\ref{metans2}) transforms to (\ref{metans1}) with
\be
f(r)=r^2 h(r^{-1})\,.
\ee

We now consider the Ans\"atze for the matter fields.  For the $SU(2)$ Yang-Mills, the general Ansatz for carrying magnetic flux is \cite{Bartnik:1988am}
\be
A\equiv
\left(
  \begin{array}{cc}
    A^3 & A^1 - {\rm i} A^2 \\
    A^1+{\rm i} A^2 & -A^3 \\
  \end{array}
\right)=
\psi\tau_1\, d\theta+(\psi\sin\theta\,\tau_2+\cos{\theta}\,\tau_3)d\varphi\,.\label{yangmills}
\ee
where $\tau_i$'s are the Pauli matrices and the function $\psi$ depends on the coordinate $r$ only.  In general the Yang-Mills field can have electric component as well, i.e.~$\phi\tau_3 dt$.  However, we fail to find any new exact solution and hence we shall not consider it here.  In the appendix, we present the equations of motion for general Ans\"atze involving the electric component as well, for more general classes of topologies, including 2-torus and hyperbolic 2-space in addition to the 2-sphere we consider in the main body of the paper.

The Maxwell equation, on the other hand, can be solved directly, given by
\be
\mathcal{A} = q \rho\, dt + p \sin\theta\, d\varphi\,.
\ee
Since the electric and magnetic charges $q$ and $p$ enter the metric only through the combination $q^2 + p^2$, we shall set $p=0$ for simplicity.

Substituting the Ans\"atze into the equations of motion, we find two independent non-linear differential equations.  The equations are significantly simpler in the $\rho$ coordinate \cite{Lu:2013hx}.  We find
\bea
&&h\psi''+h'\psi'+\psi-\psi^3 =0\,,\label{eom1}\\
&&h'h^{(3)}-\frac{h''^2}{2} +2 +\frac{1}{g_s^2} (6h\psi'^2+6\psi^2-
3\psi^4-3)-\frac{3q^2}{e^2} = 0\,.\label{eom2}
\eea
Here a prime denotes a derivative with respect to $\rho$, and $h^{(n)}$ denotes $n$ derivatives of $h$ with respect to $\rho$.  We can use (\ref{eom1}) and (\ref{eom2}) to derive that
\be h^{(4)}=\frac{6\psi'^2}{g_s^2}\,.\label{eom3}
\ee
Thus we see that although the equations contain higher derivatives, they are actually simpler than those of Einstein-Yang-Mills gravity since they involve one less function owing to the local conformal symmetry.  We shall revisit the colored black holes of Einstein-Yang-Mills theory in section 5.

It is worth commenting that the simplest solutions to (\ref{eom1}) are $\psi=0$ and $\psi=\pm 1$. When $\psi=\pm1$, the $SU(2)$ potentials become pure gauge; it is equivalent to turning off the Yang-Mills fields.  When $\psi=0$, the $SU(2)$ gauge fields are reduced to a Maxwell field (carrying magnetic charge) associated with the $U(1)$ Cartan subgroup of $SU(2)$.

\subsection{The structure of general solutions}

We do not expect to find the analytical expression of the most general solutions to (\ref{eom1}) and (\ref{eom2}).  Many properties of the solutions can nevertheless be discussed. The asymptotic expansion of the general solution can be obtained by considering the small-$\rho$ Taylor expansion, namely
\be
\psi = b_0 + b_1 \rho + b_2 \rho^2 + b_3 \rho^3 + \cdots\,,\qquad
h = a_0 + a_1 \rho + a_2 \rho^2 + a_3 \rho^3 + \cdots\,.
\ee
In terms of $r$ coordinate, the solution in large-$r$ expansion is given by
\be
\psi = b_0 + \fft{b_1}{r} + \fft{b_2}{r^2} + \fft{b_3}{r^3}\cdots\,,\qquad
f = a_0 r^2 + a_1\,r + a_2 + \fft{a_3}{r}+ \cdots\,.\label{genlarger}
\ee
Thus asymptotically as $r$ approaches infinity, the metric becomes the (A)dS spacetime with the cosmological constant $\Lambda=-3a_0$, which is an integration constant of the equations of motion. The parameters can be solved order by order.  The first few leading terms are given by
\bea
b_2 &=& \fft{b_0^3-b_0 - a_1 b_1}{2a_0}\,,\qquad
b_3=\fft{1}{6a_0^2}\Big(2a_1^2 b_1 - 2 a_1 b_0(b_0^2-1) + a_0(3b_0^2-2a_2-1)\Big) b_1\,.\cr
a_3 &=&\fft{a_2^2-1}{3a_1} + \fft{(b_0^2-1)^2-2a_0 b_1^2}{2a_1 g_s^2} + \fft{q^2}{2e^2 a_1}\,,\qquad a_4=\fft{b_1^2}{4g_s^2}\,.
\eea
Thus the asymptotic region of a general solution is characterised by six parameters, namely $(a_0,a_1, a_2, b_0,b_1,q)$.  For the solution to be asymptotic to the flat spacetime, we need to set both $(a_0,a_1)$ to zero, and $a_2=1$.

    The solutions near the horizon can also be studied by the power-series expansions.  Assuming that the horizon is located at $\rho=\rho_0$, the functions $\psi$ and $h$ can be expanded in power series of $(\rho-\rho_0)$, namely
\be
\psi=\tilde b_0 + \tilde b_1(\rho-\rho_0) + \tilde b_2 (\rho-\rho_0)^2 + \cdots\,,\qquad
h=\tilde a_1 (\rho-\rho_0) +\tilde a_2 (\rho-\rho_0)^2 + \cdots\,.
\ee
The equations of motion implies that
\bea
&&\tilde b_1 = \fft{\tilde b_0^3 -\tilde b_0}{\tilde a_1}\,,\qquad
\tilde b_2 = \fft{\tilde b_0 (\tilde b_0^2-1)(3\tilde b_0^3 - 2\tilde a_2-1)}{4\tilde a_1^2}\,,\cr
&&\tilde a_3 = \fft{\tilde a_2^2-1}{3\tilde a_1} + \fft{(\tilde b_0^2-1)^2}{2g_s^2 \tilde a_1} + \fft{q^2}{2e^2 \tilde a_1}\,,\qquad
\tilde a_4=\fft{\tilde b_0^2(\tilde b_0^2-1)^2}{4g_s^2 \tilde a_1^2}\,.
\eea
Thus we see that the near horizon geometry is specified by five independent parameters, $(r_0, \tilde a_1, \tilde a_2, \tilde b_0, q)$, one less parameter than the all allowed ones in asymptotic regions.  Thus, the counting of parameters implies that we can in general integrate out from the horizon to asymptotic (A)dS infinity and obtain black hole solutions with five independent parameters. Indeed, the numerical analysis integrating from horizons to asymptotic infinity was performed in \cite{Brihaye:2009hf,Brihaye:2009ef}.

The asymptotically-flat black holes are on the other hand less likely, and they require very delicate finite tuning, to avoid the excitation of the $a_0$ and $a_1$ terms.  However, since we have a sufficient number of parameters that we can adjust on the horizon, we expect that colored black holes in asymptotically-flat spacetime also exist.

We now calculate the global Yang-Mills charge. For our magnetic solution, it is given by \cite{Corichi:2000dm,Kleihaus:2002ee}
\be
P^{\rm YM}= \fft{\alpha}{4\pi g_s^2} \int_{r\rightarrow \infty}  \sqrt{F^a_{\theta\varphi}F^a_{\theta\varphi}}\, d\theta d\varphi = \fft{\alpha}{g_{s}^2} (b_0^2-1)\,.\label{gymc}
\ee
The global Yang-Mills charge vanishes when $b_0^2=1$, for which the corresponding term in $A$ is pure gauge. Interestingly, the global Yang-Mills charge depends only on $b_0$, but not $b_1$.  This implies that one may construct colored black hole with Yang-Mills hairs but without Yang-Mills charges.

\section{Exact solutions}

Although we do not find the general exact solutions involving five parameters to (\ref{eom1}) and (\ref{eom2}), we find many examples of special exact solutions involving reduced number of parameters.

\subsection{With only Maxwell charges}

Turning off the $SU(2)$ Yang-Mills fields is to set $\psi=\pm 1$, and hence
the equation (\ref{eom1}) is satisfied.  The equation (\ref{eom3}) implies that $h^{(4)}=0$, and hence
\be
h=a_0 + a_1 \rho + a_2 \rho^2 + a_3\rho^3\,.
\ee
Substituting this into (\ref{eom2}), one finds the following constraint:
\be
1-a_2^2 + 3 a_1 a_3 = \fft{3q^2}{2e^2}\,.\label{cons1}
\ee
In terms of the original $r$ coordinate defined by the black hole ansatz (\ref{metans1}), we have
\be
f=a_0 r^2 + a_1 r + a_2 + \fft{a_3}{r}\,,
\ee
together with the constraint (\ref{cons1}).  The metric is asymptotic (A)dS with the cosmological constant given by $\Lambda=-3a_0$.  This solution was first constructed in \cite{Riegert:1984zz}.  The thermodynamics of neutral and charged black holes was studied in \cite{Lu:2012xu} and \cite{Lu:2012ag,Li:2012gh} respectively.  Charged rotating black holes and their thermodynamic first law were studied in \cite{Liu:2012xn}.

\subsection{With only $SU(2)$ Yang-Mills charges}

\label{su2bh}

We now turn on the $SU(2)$ Yang-Mills field instead, while turning off the Maxwell field by setting $q=0$. We find a new solution for $g_s=1$, given by
\be
\psi=b_0+b_1 \rho\,,\quad h=a_0+a_1 \rho+a_2 \rho^2+a_3 \rho^3+\frac{b_1^2}{4}\rho^4\,, \label{sol2}
\ee
where
\be
a_0=\frac{(b_0^2-1)^2}{4b_1^2}>0\,,\quad a_1=\frac{b_0(b_0^2-1)}{b_1}\,,\quad a_2=\frac{3b_0^2-1}{2}\,,\quad a_3=b_0 b_1\,. \label{sol3}
\ee
The function $h$ can be factorized, namely
\be
h=\frac{(b_0+b_1 \rho-1)^2(b_0+b_1 \rho+1)^2}{4b_1^2}\,.
\ee
In terms of the $r$ coordinate, we find that the solution is given by
\be
f=a_0 r^2 \Big(1 + \fft{b_1}{(b_0-1) r}\Big)^2
\Big(1 + \fft{b_1}{(b_0+1)r}\Big)^2\,,\qquad \psi=b_0 + \fft{b_1}{r}\,.\label{fpsisol1}
\ee

To study the global structure, we first note that the metric is asymptotic to AdS spacetime with a cosmological constant $\Lambda=-3a_0<0$.  The solution has a curvature singularity at $r=0$.  The curvature singularity is shielded by an event horizon at some $r_0$ for which $f(r_0)=0$.  The location of the $r_0$ depends on the values of $(b_0,b_1)$. The function $f$ has two double roots
\be
r_1=-\fft{b_1}{b_0-1}\,,\qquad r_2=-\fft{b_1}{b_0+1}\,.
\ee
Without loss of generality, let us consider $b_1<0$.  For $b_0>1$, we have $r_1>r_2>0$ and the metric describes a single black hole where the spacetime on and outside of the horizon is specified by $r\ge r_1$.  For $-1<b_0<1$, we have $r_1<0<r_2$.  In this case, the same local metric can describe two black holes, sharing the same curvature singularity at $r=0$.  One black hole has an event horizon at $r=r_2$ and it is asymptotic to AdS as $r=+\infty$, whilst the other has the event horizon at $r=r_1$ and it is asymptotic to AdS as $r=-\infty$.  For $b_0<-1$, we have $r_2<r_1<0$, and hence the solution describes a single black hole with the event horizon at $r=r_2$ and the spacetime outside the horizon is $-\infty <r<r_2$.

Since all the horizons discussed above are associated with double roots, it follows that the temperature vanishes, giving rise to extremal black holes carrying Yang-Mills charges.  The entropy of the black holes can be calculated by the Wald entropy formula
\be
S=-\ft18 \int_{r=r_+} \sqrt{h} d^2 x\, \epsilon_{ab}\epsilon_{cd}
\fft{\partial L}{\partial R_{abcd}}=\ft13\pi\alpha \big(2 r_+\, f'(r_+) -r_+^2\, f''(r_+)+2\big)\,,\label{genentropy}
\ee
where $L={\cal L}/\sqrt{-g}$ and $r_+$ is the location of the horizon, which can be $r_1$, $r_2$, depending on the values of $(b_0,b_1)$. Substituting the solution into the entropy formula, we find that the entropy vanishes also.

To obtain the mass of the solution, we use the formula obtained in \cite{Lu:2012xu} for conformal gravity:
\begin{equation}
M=-\fft{\alpha}{8\pi} \int \lim_{r\rightarrow \infty} {\cal T}^{01}(\xi)\,,\qquad\mbox{with}\qquad {\cal T}^{\mu\nu} = C^{\mu\nu\rho\sigma} \nabla_\rho\xi_\sigma -
2\xi_\sigma \nabla_\rho C^{\mu\nu\rho\sigma}\,,\label{massformula1}
\end{equation}
where $\xi=\partial/\partial t$.  Using this formula, we find that the mass of the black hole vanishes also.

   Thus we have obtained an exact extremal black hole in conformal gravity carrying the $SU(2)$ Yang-Mills magnetic charges associated with two continuous parameters $(b_0,b_1)$, with vanishing temperature, entropy and mass.  However, the black hole contains non-vanishing global Yang-Mills charges defined by (\ref{gymc}).  How this special solution fits the general thermodynamic first law of the $SU(2)$-colored black holes will be discussed in section 4.

\subsection{With both Maxwell and Yang-Mills charges}

When $g_s=1$, we find an exact solution that carries both Maxwell and Yang-Mills charges. In terms of the $r$ coordinate, the solution is given by
\bea
f&=&\fft{(b_0^2-1)^2}{4b_1^2} r^2\Big(1 + \fft{b_1}{(b_0-1) r}\Big)^2
\Big(1 + \fft{b_1}{(b_0+1) r}\Big)^2 + \fft{q^2}{2e^2 b_1^2} r^2\,,\cr
\psi&=&b_0 + \fft{b_1}{r}\,,\qquad \mathcal{A} = \fft{q}{r} dt\,.\label{mymsol}
\eea
The solution is asymptotic to AdS with the cosmological constant
\be
\Lambda=-\fft{3(b_0^2-1)^2}{4b_1^2} - \fft{3q^2}{2e^2 b_1^2}\,.
\ee
The solution has a naked curvature singularity at $r=0$ unless $e^2<0$, for which case the Maxwell field becomes a ghost.

\subsection{Further colored black holes with a ghost-like Maxwell field}

If we allow the Maxwell field to be ghost-like, namely $e^2<0$, we find
an infinite number of solutions in which $g_s$ and $q$ take some discrete values.  Setting without loss of generality $e^2=-1$, we find
\be
g_s^2 = \fft{3n^2(3n+1)}{2(n+1)(4n^2-1)}\,,\qquad
q^2=\fft{2n(n+2)(n+3)}{3(3n+1)(n+1)^2}\,.
\ee
In terms of $r$ coordinate, the functions $\psi$ and $f$ are given by
\bea
\psi &=& b_0 \Big (1 + \fft{b_1}{nb_0 r}\Big)^n\,,\\
f &=& \fft{b_0^2}{b_1^2} r^2  \Big (1 + \fft{b_1}{nb_0 r}\Big)^2
\Big[\fft{n b_0^2}{3n+1}  \Big (1 + \fft{b_1}{nb_0 r}\Big)^{2n} - \fft{n}{n+1}\Big]\,.\label{gensolf}
\eea
(The $n=1$ case is contained in (\ref{mymsol}) as a special solution.) The cosmological constants of the asymptotic (A)dS spacetime is
\be
\Lambda= -\fft{3nb_0^2}{(3n+1)b_1^2}\Big(b_0^2 - \fft{3n+1}{n+1}\Big)\,.
\ee
Thus we have
\bea b_0^2>\fft{3n+1}{n+1},&&\quad \Lambda<0,\quad \mbox{asymptotic AdS}\,; \cr
b_0^2<\fft{3n+1}{n+1},&&\quad \Lambda>0,\quad \mbox{asymtotic dS} \,; \cr
b_0=\fft{3n+1}{n+1},&&\quad \Lambda=0,\quad \mbox{asymptotically locally flat}\,. \eea
In terms of the large-$r$ power series expansion, as defined in (\ref{genlarger}), we have
\bea
a_0 &=&-\ft13\Lambda\,,\qquad a_1 = \fft{2b_0((n+1)^2 b_0^2 - 3n-1)}{(n+1)(3n+1) b_1}\,,\cr
a_2&=&\fft{(2n+1)(n+1)^2 b_0^2-3n-1}{n(n+1)(3n+1)}\,,\qquad
a_3=\fft{2(n+1)(2n+1)b_0b_1}{3n(3n+1)}\,.
\eea
It is worth pointing out that we have implicitly assume that $n$ is an integer.  However, the equations of motion are satisfied for any $n$. The solutions describe black hole with the horizon located at the largest real root of the square bracket in (\ref{gensolf}).

\section{The first law of thermodynamics}

In section \ref{theory}, we considered black holes in conformal gravity carrying both $SU(2)$ Yang-Mills and $U(1)$ Maxwell charges.  We obtained the general structure of the solutions and find that the solution contain five parameters including the cosmological constant.  In section \ref{su2bh}, we obtained a specific black hole solution carrying only the $SU(2)$ Yang Mills charges with two independent parameters.  This solution is rather unusual in that it has vanishing temperature, entropy and mass.  In this section, we shall obtain the thermodynamic first law for the general black hole and study what becomes the remainder of the first law for this special solution.

One key quantity of a black hole is its mass, or energy.  For higher derivative gravities, the mass can be rather subtle.  If the asymptotic behavior of the solution takes the form
\be
f=-\ft13\Lambda r^2 + 1 - \fft{2M}{r} + \fft{a_4}{r^2} + \cdots\,,
\ee
One can use either the ADT method \cite{destek} or the generalization of \cite{Okuyama:2005fg,Pang:2011cs} the AMD method \cite{Ashtekar:1984zz,Ashtekar:1999jx} to show that the mass is $-\fft23 \alpha \Lambda M$.  The asymptotic structure of our solution however contains extra terms such as $a_1 r + a_2$ that falloff slower than the mass term, as indicated in (\ref{genlarger}).  These slower falloffs are associated with the massive spin-2 modes in higher derivative gravity and the above methods for calculating black hole mass are no longer valid for these more general solutions.  In \cite{Lu:2012xu}, the mass formula (\ref{massformula1}) was derived by considering the Noether charge associated with time-like Killing vector, based on the Wald formalism.  In this section, we adopt the Wald formalism to derive the thermodynamical first law of colored black holes.

\subsection{Deriving the first law}

The Wald formalism of Einstein gravity coupled to a Maxwell or a Proca field was previously studied in \cite{gao} and \cite{Liu:2014tra} respectively, and hence we shall not include it in this discussion.  Thus for the Lagrangian (\ref{genlag}) with ${\cal A}$ turned off, the variation allows us to obtain the equations of motion and the Noether current one-form:
\be \delta(\sqrt{-g}\mathcal{L})=\sqrt{-g}(E_{\mu\nu}\delta g^{\mu\nu} + E^a_\mu \delta A_a^\mu +\triangledown_\mu J^\mu)\,,
\ee
where $E^a_\mu=0$ and $E_{\mu\nu}=0$ are the Yang-Mills and Einstein equations of motion respectively. Here $J^\mu$ has contributions from both gravity and the Yang-Mills field: $J^\mu = J^\mu_{(G)}+J^\mu_{(A)}$. To be specific, we have
\bea
J^{\mu}_{(G)}&=&(\ft 23R G^{\mu\nu\rho\lambda}-2 T^{\mu\nu\rho\lambda}+4 R^{\mu\rho\lambda\nu})\nabla_\nu \delta g_{\rho\lambda} \cr
&&-(\ft23 G^{\mu\nu\rho\lambda}\nabla_\nu R-2 \nabla_\nu T^{\nu\mu\rho\lambda}+4 \nabla_\nu R^{\mu\rho\lambda\nu}) \delta g_{\rho\lambda}\,,\cr
J^\mu_{(A)} &=& -\ft{2\alpha}{g_s^2}F^{a\mu\nu}\delta A_\nu^a\,,
\eea
where
\be G^{\mu\nu\rho\sigma}=\ft12 (g^{\mu\rho}g^{\nu\sigma}+g^{\mu\sigma}g^{\nu\rho})-g^{\mu\nu}g^{\rho\sigma},
\quad T^{\mu\nu\rho\lambda}=g^{\mu\rho}R^{\nu\lambda}+
g^{\mu\lambda}R^{\nu\rho}-g^{\mu\nu}R^{\rho\lambda}-g^{\rho\lambda}R^{\mu\nu}\,. \ee
From the current $J^\mu$, one can define the Noether current 1-form and its 3-form Hodge dual as:
\be J_{\1}=J_{\mu}dx^\mu,\qquad \Theta_{\3}={*J_{\1}},\qquad J_{(3)}=\Theta_{(3)}-i_{\xi}\cdot {*\mathcal{L}}\,,
\ee
where $\xi$ is a Killing vector and $i_\xi\cdot$ denotes that $\xi$ contracts with the first index of the tensor it acts. It is fairly straightforward to verify that for a diffeomorphism invariant Lagrangian, one always has $dJ_{(3)}= 0$, provided that the equations of motion are satisfied.  Thus one can further define a Noether charge 2-form, $J_{\3}=dQ_{\2}$.  It was shown in \cite{wald1,wald2} that for a given gravitational solution, the variation of the Hamiltonian with respect to the integration constants is given by
\be
\delta H=\int_{\mathcal{C}}\delta J_{\3}-d(i_{\xi}\cdot \Theta_{\3})=\int_{\Sigma_2}\delta Q_{\2}-i_{\xi}\cdot \Theta_{\3}\,,
\ee
where $\xi$ is the time-like Killing vector that becomes null on the event horizon. The first law of thermodynamics of black holes in any gravity theories can be derived from the Wald formula $\delta H=\delta H_\infty- \delta H_{+}$ evaluated on the horizon and at the asymptotic infinity respectively. For our conformal gravity with the Yang-Mills field, we have
\bea Q_{\2}^{(G)} &=& -\frac{\alpha}{16\pi}\varepsilon_{\mu\nu\lambda\tau} (C^{\mu\nu\rho\sigma}\nabla_\rho \xi_\sigma-2\xi_\sigma
\nabla_\rho C^{\mu\nu\rho\sigma})\,,\cr
Q_{\2}^{(A)} &=& -\fft{\alpha}{16\pi g_s^2}
\varepsilon_{\mu\nu\lambda\tau}F^{a\mu\nu}
A^a_{\sigma}\xi^{\sigma}\,.
\eea
The total charge 2-form is given by the sum of the above two contributions.  Note that in presenting $\delta H$, we multiply our Lagrangian (\ref{genlag}) by an overall $1/(16\pi)$.

For our Ans\"atze (\ref{metans1}) and (\ref{yangmills}), it is immediately clear that $Q_{(2)}^{(A)}=0$.  Therefore, we have
\bea
&& Q_{\2} = \frac{\alpha\Omega_\2}{48\pi}\Big(k(-\frac{4f}{r}+2f')
+\frac{4f^2}{r}-6ff'+2r f'^2+2r ff''-r^2f'f''+2r^2ff^{(3)}\Big),\cr
&&i_\xi\cdot \Theta_{(3)}^{(G)}=\frac{\alpha\Omega_\2}{48\pi} \Big(k(-\frac{4\delta f}{r}+2\delta f')+\delta f (\frac{4f}{r}-4f'+2r f''+r^2f^{(3)})\cr
&&\qquad\qquad\qquad+\delta f'(-2f+2r f'-r^2f'')\Big)\,,\cr
&&i_\xi\cdot \Theta_{(3)}^{(A)}=\frac{\alpha\Omega_\2}{4\pi g_s^2}f \psi'\delta\psi\,,
\eea
where $\Omega_\2$ is the volume 2-form of the 2 dimensional space of $(\theta,\varphi)$. In presenting the result, we introduced a topological parameter $k=1,0,-1$, for the formulae to be applicable for topologies of torus and hyperbolic spaces as well.  (See appendix A for a discussion of colored black holes in different topologies.) Given the above results, we can deduce that
\bea
\delta Q_{\2}-i_{\xi}\cdot \Theta_{\3} &=& \frac{\alpha\Omega_\2}{48\pi} \Big(\delta f (\frac{4f}{r}-2f'+r^2f^{(3)})+\delta f'(-4f+2r f')\cr
&&+\delta f''(2r f-r^2 f')+\delta f^{(3)}2r^2 f-\frac{12}{g_s^2}f\psi'\delta \psi\Big)\,.
\eea
Interestingly, this result is independent of the topological parameter $k$.  Note that in this section, a prime denotes a derivative with respect to $r$, rather than $\rho$ in section 2.  Analogous formulae for black holes with spherical/toric/hyperbolic isometries in various gravity theories coupled to scalar and vector fields were obtained in \cite{Liu:2013gja,Liu:2014tra,Lu:2014maa,Liu:2014dva}.

For the large-$r$ expansion, $\psi$ and $f$ take the forms given in (\ref{genlarger}).  We can read off the mass of the solution using the Noether charge 2-form,
\be
M=\int_{r\rightarrow \infty} Q_\2= \ft1{6} \alpha \Big(a_1 (a_2-k) - 6a_0 a_3\Big)\fft{\omega_2}{4\pi}\,.\label{massformula2}
\ee
where $\omega_2=\int \Omega_\2$ is the volume for the foliating 2-space. For $S^2$, we have $\omega_2=4\pi$.  Note that the mass formula (\ref{massformula2}) is equivalent to the one defined in (\ref{massformula1}). It is easy to verify that the mass for the Schwarzschild-AdS black hole with $f=a_0 r^2 + 1-2m/r$, which is a solution of conformal gravity, is given by
\be
M=2\alpha a_0 m\,,
\ee
which was also obtained in \cite{Lu:2012xu}.  It should be emphasized that it is rather special a property of conformal gravity that the quantity $\int_{r\rightarrow\infty} Q_\2$ is finite.  This term is typically divergent for asymptotic AdS black holes in Einstein gravity with a cosmological constant and it requires $i_\xi\cdot \Theta$ to cancel the divergence.

It is easy to verify that evaluating the $\delta H$ near the horizon geometry gives rise to
\be
\delta H_+ = T \delta S\,,
\ee
where
\be
T=f'(r_+)/(4\pi)\,,\qquad
S=\ft13\pi\alpha \big(2 r_+\, f'(r_+) -r_+^2\, f''(r_+) + 2k\big)\fft{\omega_2}{4\pi}\,.\label{tempentr}
\ee
Evaluating $\delta H$ at the asymptotic infinity gives
\bea
\delta H_\infty  &=& \ft16 \alpha \Big(-3 a_3 \delta a_0 - 6 a_0 \delta a_3
+ a_1\delta a_2 + 6 g_s^{-2} a_0 b_1 \delta b_0\Big)\fft{\omega_2}{4\pi}\cr
&=& \delta M - \ft1{6} (a_2-k)\alpha \delta a_1 + \ft12 \alpha a_3 \delta a_0 +
\alpha g_s^{-2} a_0 b_1 \delta b_0\,.
\eea
We follow (\ref{gymc}) and also the discussion in \cite{Lu:2012xu}, and define the massive spin-2 hair, Yang-Mills charge and thermodynamical pressure and their thermodynamical conjugates as follows:
\bea
 \mbox{massive spin two hair:}&& \Xi=\fft{\alpha\omega_2}{4\pi}\,a_1,\qquad \Phi=\ft1{6}(a_2-k)\,,\cr
\mbox{Yang-Mills hair:} && P^{\rm YM}=\fft{\alpha\omega_2}{4\pi g_s^2}\,(b_0^2-1),\qquad \Phi^{\rm YM}= -\fft{a_0 b_1}{2b_0}\,,\cr
\mbox{thermodynamic pressure:}&& \Lambda=-3a_0,\qquad V=\frac{\alpha\omega_2 a_3}{24\pi }\,.
\eea
We thus deduce that the thermodynamical first law of general black holes carrying the magnetic $SU(2)$ Yang-Mills charges are given by
\be
dM=TdS+\Phi d\Xi+Vd\Lambda+\Phi^{\rm YM} dP^{\rm YM}\,.\label{firstlaw}
\ee
If we also consider solutions with electric charges of the Maxwell field, the first law becomes
\be
dM=TdS+\Phi d\Xi+Vd\Lambda+\Phi^{\rm YM} dP^{\rm YM} + \Phi_e dQ\,,\label{firstlaw1}
\ee
where $\Phi_e$ is the electric potential and $Q$ electric charge
\be
Q=\fft{\alpha}{8\pi e^2} \int {*{\cal F}} = \fft{\alpha}{2e^2} q\,.
\ee
Thus the five integration constants of the general solution considered in section 2 correspond to the mass $M$, electric and Yang-Mills charges $(Q, P^{\rm YM})$, massive spin-2 hair $\Xi$ and thermodynamical pressure associated with the cosmological constant $\Lambda$.  In \cite{Kastor:2009wy,Cvetic:2010jb}, the cosmological constant in two-derivative gravity was treated as the thermodynamical pressure and its conjugate as the volume was derived from generalizing the usual first law by treating the cosmological constant as a variable rather than a fixed constant.  Since the cosmological constant appears in the action in that case, this generalized first law cannot be derived from the Wald formalism.  In our case, the cosmological constant arises naturally as an integration constant and hence its contribution to the first law arises naturally via the Wald formalism.

\subsection{Testing the first law}

Having obtained the first law and the formulae for various thermodynamical quantities, we can now test it using the explicit solutions we obtained in section 3.  The first example we would like to examine is the extremal black hole solution obtained in section 3.2. It is easy to verify that for this solution we have indeed vanishing mass, temperature and entropy, implying that the first law reduces to
\be
\Phi d\Xi + P d\Lambda + \Phi^{\rm YM} dP^{\rm YM} = 0\,.\label{ymfl}
\ee
We now verify that it is indeed the case. For $f$ and $\psi$ given in (\ref{fpsisol1}), we have
\bea
&&\Phi=\ft18 (b_0^2-1)\,,\qquad \Xi=\fft{\alpha b_0^2 (b_0^2-1)^2}{b_1^2}\,,\qquad
V=\ft16\alpha b_0^2(b_0^2-1)\,,\cr
&&\Lambda=-\fft{3 (b_0^2-1)^2}{4 b_1^2}\,,\qquad P^{\rm YM}=\alpha (b_0^2-1)\,,\qquad
\Phi^{\rm YM}=-\fft{(b_0^2-1)^3}{8b_1^2}\,.
\eea
It is then straightforward to verify that the remainder of the first law (\ref{ymfl}) is satisfied by the black hole solution.

Next, we consider the general class of solutions obtained in section 3.4.  For these solutions we treat the parameter $n$ as fixed constant taking integer values, and hence the electric charge does not involve in the first law. From its asymptotic behavior, we can determine
\bea
M &=&\fft{\alpha b_0}{3 n (n+1)^2 (3 n+1)^2 b_1}\Big(
(n^2+n+1)(3n+1)^2 \cr &&
\qquad+ (n-1)(n+1)^2(3n+1)(3n+2) b_0^2 -(n-1)(n+1)^3 (2n+1) b_0^4\Big)\,,\cr
\Xi &=& \fft{2\alpha b_0 ((n+1)^2 b_0^2 - 3n-1)}{(n+1)(3n+1) b_1}\,,\quad
\Phi = \fft{(n+1)^2(2n+1)b_0^2-3n-1}{6n(n+1)(3n+1)} - \fft16\,,\cr
\Lambda &=& - \fft{3nb_0^2 ((n+1)b_0^2-3n-1)}{(n+1)(3n+1)b_1^2}\,,\qquad
V=\fft{(n+1)(2n+1)\alpha b_0 b_1}{9n(3n+1)}\,,\cr
P^{\rm YM} &=& \fft{2(n+1)(4n^2-1)\alpha}{3n^2(3n+1)} (b_0^2-1)\,,\qquad
\Phi^{\rm YM}=\fft{n(3n+1-(n+1)b_0^2)b_0}{2(n+1)(3n+1)b_1}\,.
\eea
Together with (\ref{tempentr}), it is then straightforward to verify that the first law (\ref{firstlaw}) is satisfied by these black hole solutions.  The double root in the expression of $f$ in (\ref{gensolf}) might suggest that the solution is extremal with zero temperature.  However, the horizon is always located at the root for which the square bracket term in (\ref{gensolf}) vanishes, and hence the solution has non-vanishing temperature.  Interestingly, it turns out that the entropy is always a pure numerical constant, namely
\be
S=-\fft{2\alpha (n+2) \pi}{3(n+1)}\,.
\ee
Thus the term $TdS$ vanishes, even though $T\ne 0$.  However, this does not imply that the usual specific heat vanishes in our system.  In fact for our solutions with only two variables $(b_0,b_1)$, we cannot define the specific heat $C=T(\partial S/\partial T)$ while holding all $(\Xi, \Lambda, P^{\rm YM})$ fixed.  Note also that since $S$ is independent of the variables $(b_0,b_1)$, it follows that we can always remove the $S$ by adding the Gauss-Bonnet term in the action with some appropriate coefficient.  (The Gauss-Bonnet term in the action contributes a fixed constant to the entropy of a black hole.) It should be emphasized that it is only in higher-derivative gravity where the entropy area law no longer applies that it is possible to have a black hole with smooth horizon but fixed constant entropy.

Finally we consider the solution (\ref{mymsol}) that carries both the continuous Maxwell and Yang-Mills charges.  For this solution to describe a black hole, we need to let $e^2<0$, in which case the Maxwell field is ghost-like. Without loss of generality, we set $e^2=-1$, and find
\bea
&&M=\fft{\alpha b_0 q^2}{2 b_1}\,,\qquad
\Xi=\fft{\alpha b_0 (b_0^2-1)}{b_1}\,,\qquad \Phi=\ft14(b_0^2-1)\,,\cr
&&P^{\rm YM}=\alpha (b_0^2-1)\,,\qquad\Phi^{\rm YM}= -\fft{3(b_0^2-1)^2-6q^2}{24b_0 b_1}\,,
\qquad \Lambda=-\fft{3(b_0^2-1)^2-6q^2}{4 b_1^2}\,,\cr
&&V=\ft16\alpha b_0 b_1\,,\qquad Q=-\ft12\alpha q\,,\qquad \Phi_e=\fft{q}{r_0}\,.
\eea
Together with the temperature and entropy (\ref{tempentr}), it is easy to verify that the first law (\ref{firstlaw1}) is satisfied.

\section{$SU(2)$-colored black holes in Einstein gravity revisited}

Having studied the colored black holes in conformal gravity and their first law of thermodynamics, we would like to revisit the same issue in Einstein gravity. We consider the Lagrangian
\be
\mathcal{L}=\sqrt{-g} \Big(R-2\Lambda - \frac{1}{2g_s^2}F^a_{\mu\nu}F^{a\mu\nu}\Big)\,.\label{eymsu2}
\ee
The $SU(2)$-colored black hole ansatz is given by \cite{Bartnik:1988am}
\bea
ds^2 &=& -fe^{-2\chi}dt^2+\frac{dr^2}{f}+r^2(d\theta^2 + \sin^2\theta d\varphi^2)\,,\cr
A &=&\psi\tau_1 d\theta+(\psi\tau_2+\cot{\theta}\tau_3)\sin{\theta}d\varphi\,.
\eea
The equations of motion for the three functions $f(r), \chi(r)$ and $\psi(r)$ are given by
\bea
&&r^2f\psi''+ r^2(f' - f \chi') \psi' - \psi^3 + \psi=0\,,\qquad
r \chi' + \fft{\psi'^2}{g_s^2}=0\,,\cr
&& r^2 (-1 + r^2 \Lambda + f + r f') + \fft{1}{4g_s^2} (2r^2 f \psi'^2 +
(\psi^2-1)^2)=0\,.
\eea
One way to examine whether there are asymptotic AdS black holes is to study the linearized Yang-Mills modes in the AdS background, corresponding to $f=r^2/\ell^2 +1$, $\chi=0$ and $\psi^2=1$, where $\Lambda=-3/\ell^2$.  Performing small perturbation of $\psi$, namely $\psi=\pm 1 + \hat \psi$,  at the linear level, we find
\be
\hat \psi=\fft{\ell}{r}\Big(\hat b_0 (\ell^{-1}r - \arctan(\ell^{-1}r)) + \hat b_1\Big)\,.
\ee
It is clear that both modes are convergent at large $r$ and their back reaction to gravity will preserve the asymptotic AdS structure, at least at the linear level.  This is a strong indication that colored black holes with global Yang-Mills charges exist.  To study the structure of these black holes,  we consider the large $r$ expansion, for which we find that
\bea
\psi &=& b_0 + \fft{b_1}{r} -\fft{3b_0(b_0^2-1)}{2\Lambda r^2} -
\fft{3b_1(b_0^2-1)}{2\Lambda r^3} + \cdots\,,\qquad
\chi = c_0 + \fft{b_1^2}{4g_s^2 r^4}+ \cdots\,,\cr
f &=& -\ft13 \Lambda r^2 + 1 - \fft{2M}{r} +\fft1{6g_s^2 r^2} \big(3(b_0^2-1)^2 -
2\Lambda b_1^2\big) + \cdots\,.\label{einslarger}
\eea
Thus we see that there are four independent integration constants in the asymptotic AdS expansion, namely $(M, b_0, b_1,c_0)$. The constant $c_0$ should be set to zero so that $t$ is the proper time in asymptotic AdS spacetimes.  Assuming there exists a black hole whose horizon is located at $r=r_0$, the power-series expansion of the functions near the horizon geometry is given by
\bea
\psi &=& \tilde b_0 + \tilde b_1(r-r_0) + \tilde b_2(r-r_0)^2 + \cdots\,,\cr
\chi &=& \tilde c_0 + \tilde c_1 (r-r_0) + \tilde c_2 (r-r_0)^2 + \cdots\,,\cr
f &=& \tilde a_1(r-r_0) + \tilde a_2 (r-r_0)^2 + \cdots\,.
\eea
Substituting these into the equations of motion and solve them order by order, we find that for the near horizon expansion, there are three parameters $(\tilde b_0,\tilde c_0, r_0)$. The parameter $\tilde c_0$ however is trivial in that it has to be adjusted so that $c_0=1$ at the asymptotic infinity when the solution is integrated out to infinity.  Thus the near horizon geometry is specified by two independent non-trivial parameters.  Since the asymptotic AdS expansion contain all the integration constants, it follows that the near-horizon geometry can in general be integrated out to the asymptotic infinity where the parameters $(M,b_0,b_1)$ are subject to one algebraic constraint.  Thus the colored AdS black holes exist and contain two independent parameters.  The numerical analysis to establish the existence of these solutions were performed in \cite{Torii:1995wv,Winstanley:1998sn}.

    Our work in conformal gravity in earlier sections shows that one does not need to know the full exact solution in order to derive the first law.  The solution's existence and the asymptotic expansions to some appropriate orders are enough. To derive the thermodynamical first law of the colored AdS black holes, we obtain the reduced Wald formula in this special case, namely
\be
\delta H=-\frac{\omega_2}{16\pi}e^{-\chi}(2r\delta f+\frac{4}{g_s^2}f\psi'\delta \psi)\,.
\ee
It is clear that $\delta H_+ = T dS$ and
\be
\delta H_\infty=\delta M - \fft{b_1\Lambda}{3g_s^2} \delta b_0\,.
\ee
This give rise to the first law
\be
dM= T dS +\Phi^{\rm YM} d P^{\rm YM}\,,
\ee
where $P^{\rm YM}=(b_0^2-1)/g_s^2$ is the $SU(2)$ Yang-Mills magnetic charge and $\Phi^{\rm YM}=\Lambda b_1/(6b_0)$ is the conjugate potential.

The situation is quite different in asymptotic Minkowski spacetimes with $\Lambda=0$.  The linear expansion of the Yang-Mills field gives rise to
\be
\hat \psi=\hat b_0 r^2 + \fft{\hat b_1}{r}\,.
\ee
Thus we see that one of the mode, namely the $\hat b_0$ mode is divergent, and its back reaction to gravity will break the Minkowski spacetime.  Thus this mode has to be excluded in the solution, implying the absence of any global Yang-Mills charge.  Indeed, in the large $r$ expansion (\ref{einslarger}), when $\Lambda=0$, the parameter $b_0$ is forced to $\pm1$, corresponding to the pure gauge of the Yang-Mills fields. Thus, the asymptotically-flat colored black hole is unlikely to exist since it is unlikely not to excite the $b_0$ mode when integrated out from the horizon.

It was however found numerically in \cite{Bizon:1990sr} that there can exist colored black holes in Einstein-Yang-Mills gravity with vanishing cosmological constant, through delicate fine tuning.  For these back holes, there is only one continuous variable $M$, with $b_0=0$ and $b_1$ being a function of mass $M$.
It is then clear that the first law is simply
\be
dM = TdS\,.
\ee
In other words, these asymptotically-flat black holes contain $SU(2)$ hair, but no $SU(2)$ charges, and first law takes the same form as that of Schwarzschild black hole.

It is worth remarking that the Lagrangian (\ref{eymsu2}) can be lifted to $D=11$ supergravity when $\Lambda=\pm 4g_s^2$ \cite{Pope:1985bu,Lu:2003dm}, i.e.~the cosmological constant can be both positive and negative.  The lifting of the non-abelian black holes to $D=11$ was discussed in \cite{Mann:2006jc}.

\section{Conclusions}

In this paper, we considered conformal gravity coupled to the non-abelian $SU(2)$ Yang-Mills field and the abelian $U(1)$ Maxwell field.  We studied spherically-symmetric and static black holes carrying electric $U(1)$ and magnetic $SU(2)$ charges.  We found that the general solutions contain five parameters, describing the mass, the $U(1)$ and $SU(2)$ charges, the massive spin-2 charge and the cosmological constant.  We adopted the Wald formalism to derive the first law of thermodynamics of these black holes.  We obtained a few examples of exact solutions carrying $SU(2)$ charges.  We verified that the general first law were indeed satisfied by these special solutions.

    The exact $SU(2)$-colored black holes we obtained have some very unusual properties.  One of them has zero mass, temperature and entropy, but it has non-vanishing global $SU(2)$ Yang-Mills charges.  We obtained the remainder of the general first law for this special solution.  We also find a class of $SU(2)$-colored black holes carrying fixed $U(1)$ charges of the ghost-like Maxwell field.  Interestingly, these black holes with regular event horizons can have vanishing entropy but non-zero temperature.

    The exact solutions of $SU(2)$-colored (A)dS black holes that we have obtained in this paper are the first examples of the kind in literature.  They may provide us new toys for studying both higher-derivative gravities and colored black hole physics.

\section*{Acknowledgement}

We are grateful to Hai-Shan Liu for useful discussions.  Z.-Y.~Fan is supported in part by NSFC Grants NO.10975016, NO.11235003 and NCET-12-0054; The work of H.L.~is supported in part by NSFC grants NO.11175269, NO.11475024 and NO.11235003.

\appendix

\section{Colored topological black holes with Yang-Mills charges }

In this appendix, we consider $SU(2)$-colored black holes in different topologies, including torus and hyperbolic 2-spaces, in addition to 2-sphere studied in the main body of the paper.  In terms of the $r$ coordinate, the metric ansatz is given by
\be
ds^2=-f(r)dt^2+\frac{dr^2}{f(r)}+r^2\Big(\frac{dx^2}{1-k x^2}+(1-k x^2)dy^2\Big)
\,,
\ee
where $k=-1,0,1$ and the corresponding foliating 2-space of $(x,y)$ is a 2-sphere, 2-torus and hyperbolic 2-space respectively.  In terms of the
$\rho=1/r$ coordinate, as discussed in section 2, the metric can be conformally transformed to
\be
ds^2=-h(\rho)dt^2+\frac{dr^2}{h(\rho)}+\frac{dx^2}{1-k x^2}+(1-k x^2)dy^2
\,,
\ee
where $h(\rho)=\rho^2 f(1/\rho)$.  The $U(1)$ and $SU(2)$ Yang-Mills potentials are then given by
\be
{\cal A}=q\rho dt\,,\qquad A=\phi\tau_3 dt+\frac{\psi}{\sqrt{1-k x^2}}\tau_1 dx+(\psi\sqrt{1-k x^2}\tau_2-k x\tau_3)dy\,,
\ee
where $q$ is a constant and $\phi$ and $\psi$ are functions of $\rho$.  Note that here for completeness, we have also introduced the electric potential $\phi$ for the $SU(2)$ Yang Mills field that we did not consider in the main body of the paper.  The full set of equations of motion are then reduced to the three different equations
\bea
&&\phi''-\frac{2\psi^2}{h}\phi=0\,,\qquad h\psi''+h'\psi'+(k+\frac{\phi^2}{h})\psi-\psi^3=0\,,\cr
&&h'h^{(3)}-\frac{h''^2}{2} +\frac{6h\psi'^2}{g_s^2}+\frac{6}{g_s^2}(k+\frac{\phi^2}{h})\psi^2
-\frac{3(\psi^4+\phi'^2)}{g_s^2}+k^2(2-\frac{3}{g_s^2})-\frac{3q^2}{e^2}=0\,.
\eea
Note that in this appendix, a prime denotes a derivative with respect to $\rho$.  These equations imply that
\be
f^{(4)}=\frac{6}{g_s^2}(\psi'^2+\frac{\phi^2\psi^2}{f^2})\,.
\ee
We shall not attempt to solve for the general equations here, but present only one special exact solution with $\phi=0$ and $g_s=1$.  In terms of $r$ variables, we find
\be
\psi=b_0 + \fft{b_1}{r}\,,\qquad
f=r^2\Big(\frac{[(b_0+b_1/r)^2-k]^2}{4b_1^2}+\frac{q^2}{2e^2 b_1^2}\Big)\,.
\ee
The $k=1$ solution was already presented in section 3, which can describe a black hole when $q=0$.  It ceases to become a black hole for $k=0$ or $-1$.

Finally note that in terms of $(\theta,\varphi)$ coordinates, defined by
\be
x=-\frac{\cos{(\sqrt{k}\theta)}}{\sqrt{k}},\quad y=\frac{\varphi}{\sqrt{k}}\,,
\ee
The full Ans\"atze can be written as
\bea
ds^2 &=&-f(r)dt^2+\frac{dr^2}{f(r)}+r^2
(d\theta^2+\frac{\sin^2{(\sqrt{k}\theta)}}{k}d\varphi^2)\,,\cr
A&=&\phi\tau_3 dt+\psi\tau_1 d\theta+ (\psi\tau_2+\cot{(\sqrt{k}\theta)}\tau_3) \sin{(\sqrt{k}\theta)}d\varphi\,,\qquad \mathcal{A}=\fft{q}{r} dt\,,
\eea

\end{document}